\documentclass[letter]{aa}
\usepackage{graphicx}
\usepackage{txfonts}
\usepackage[colorlinks=true,citecolor=blue]{hyperref}
\begin{document} 

   \title{The disk of 2MASS~15491331-3539118 = GQ~Lup~C as seen by HST and WISE\thanks{Based on observations made with the NASA/ESA Hubble Space Telescope, obtained from the data archive at the Space Telescope Science Institute. STScI is operated by the Association of Universities for Research in Astronomy, Inc. under NASA contract NAS 5-26555}}

   \author{C. Lazzoni\inst{1,2}
          \and
           R. Gratton\inst{1}
          \and
           J. M. Alcal\'a\inst{3}
          \and
          S. Desidera\inst{1}
          \and
          A. Frasca\inst{4}
          \and
          C.F. Manara\inst{5}
          \and
          D. Mesa\inst{1}
          \and
          E.  Rigliaco\inst{1}
          \and
          A. Vigan\inst{6}
          \and
          A. Zurlo\inst{7,8}
}

   \institute{INAF -- Osservatorio Astronomico di Padova, Vicolo dell'Osservatorio 5, I-35122, Padova, Italy
              \email{cecilia.lazzoni@inaf.it}
              \and 
              Dipartimento di Fisica a Astronomia "G. Galilei", Universita' di Padova, Via Marzolo, 8, 35121 Padova, Italy
              \and
              INAF -- Osservatorio Astronomico di Capodimonte, Salita Moiariello 16, 80131 Napoli, Italy
              \and
              INAF -- Osservatorio Astronomico di Catania, Via Santa Sofia, 78, Catania, Italy
              \and
              European Southern Observatory, Karl-Schwarzschild-Strasse 2, 85748, Garching bei München, Germany
              \and 
              Aix Marseille Universit\'e, CNRS, LAM (Laboratoire d'Astrophysique de Marseille) UMR 7326, 13388 Marseille, France
    \and
    N\'ucleo de Astronom\'ia, Facultad de Ingenier\'ia y Ciencias, Universidad Diego Portales, Av. Ejercito 441, Santiago, Chile 
    \and
    Escuela de Ingenier\'ia Industrial, Facultad de Ingenier\'ia y Ciencias, Universidad Diego Portales, Av. Ejercito 441, Santiago, Chile 
            }

   \date{Received ; accepted }

 
  \abstract
   {}
   {Very recently, a second companion on wider orbit has been discovered around GQ~Lup. This is a low-mass accreting star partially obscured by a disk seen at high inclination. If detected, this disk may be compared to the known disk around the primary.}
   {We detected this disk on archive HST and WISE data.}
   {The extended spectral energy distribution provided by these data confirms the presence of accretion from H$\alpha$ emission and UV excess, and shows an IR excess attributable to a warm disk. In addition, we resolved the disk on the HST images. This is found to be roughly aligned with the disk of the primary. Both of them are roughly aligned with the Lupus I dust filament containing GQ Lup. 
   }
   {}


   \maketitle
%

\section{Introduction}

\citet{Alcala2020} (hereinafter Paper I) recently found a probable second companion on wider orbit (2MASS~15491331-3539118; projected separation $\sim$2400~au) to the very young star GQ~Lup using Gaia DR2 data \citep{Gaia2018}, that also has a closer Brown Dwarf (BD) companion (projected separation $\sim$100 au). Paper I found that GQ~Lup~C is accreting; their data strongly suggests that it may be surrounded by a disk seen nearly edge-on that attenuates the star light, while possibly leaving unperturbed emission by an outflow. However, the disk could not be detected from the spectral energy distribution (SED) considered in that paper, that only extends up to the K-band. The purpose of this paper is to find further evidences for this disk. We found that this can be obtained both extending the spectral range, by considering observations acquired with WISE, and using high spatial resolution images such as those provided by HST. On the other side, GQ~Lup~C is too far from the primary to have been observed with the previous survey with ALMA.

\section{Archival data}

\subsection{HST data}

Images of the region around GQ Lup in nine different bands were obtained with the Hubble Space Telescope (HST) Wide Field Camera 3 (WFC3) (Proposal 12507: Kraus et al.) at epoch 2012.15. These images include GQ~Lup~C. We retrieved them using the MAST archive interface (https://archive.stsci.edu/hst/) in order both to extend the SED to the UV and to exploit the high space resolution to possibly resolve the disk. We also retrieved HST NICMOS images but GQ~Lup~C resulted to be outside the observed field.

\subsection{WISE data}

\begin{figure}
   \centering
   \includegraphics[width=\columnwidth]{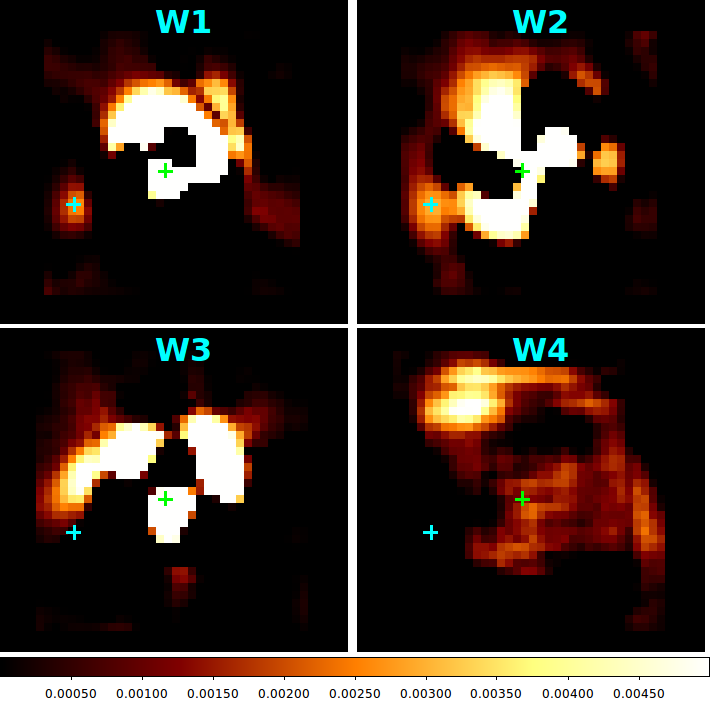}
   \caption{Images in the four different bands of WISE after subtraction of the signal from GQ Lup A, as described in the text. The position of GQ Lup A and C are marked by green and cyan crosses, respectively. In all these images, North is up and East to the left. GQ~Lup~C is detected in W1 and W2, but not at longer wavelengths}
   \label{fig:wise}
   \end{figure}

GQ~Lup~C is not listed in the WISE \citep{Wright2010} point source catalogue because it is not well resolved from the much brighter primary. In order to constrain the thermal infrared emission from GQ~Lup~C, we then retrieved WISE images of the region around GQ Lup using the NASA-IRSA interface (https://irsa.ipac.caltech.edu/Missions/wise.html). As mentioned above, the WISE Point Spread Function (PSF) is very extended (Full With at Half Maximum FWHM$\sim 10$~arcsec, that is much larger than the size of the disk around A) and C is much fainter than A. To reduce the contamination by A, we simply subtracted from each image the same image rotated by 180 degree. The exact position of the pivot of the rotation was obtained by minimizing the r.m.s. within a square region of 11$\times $11 pixels centered on a first approximate position obtained by a Gaussian fitting routine. In this way, symmetric features on the PSF are suppressed\footnote{We also downloaded images of nearby objects, but their images look quite different so that cannot be used for a PSF subtraction}. Figure~\ref{fig:wise} gives the result of this subtraction. There is a clear signal at the position of C in W1 and W2, while nothing is detected in W3 and W4. We however used the non-detection in W3 to provide an upper limit to the emission by GQ~Lup~C, while we did an aperture photometry for W1 and W2. 

\section{Spectral Energy Distribution}

   \begin{figure}
   \centering
   \includegraphics[width=\columnwidth]{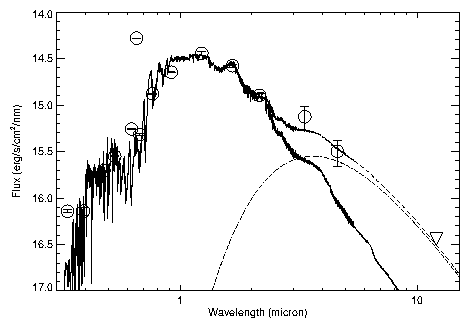}
   \caption{Observed spectral energy distribution for GQ~Lup~C; circles are measures form HST, 2MASS and WISE data; the down-pointed triangle is the upper limit in the W3 band. The solid line is the prediction for a BT-Settl model \citep{Allard2012} of 3200~K, with an absorption of $A_V=1.0$~mag (see Paper~I), the dash-dotted line is a black body spectrum with a temperature of 800~K, and the dashed line is the sum of the two. The discordant point is the HST-F656N H$\alpha$ band }
   \label{fig:sed}
   \end{figure}

In order to discuss the spectral energy distribution (SED) of GQ~Lup~C, we list in Table~\ref{tab:sed} the fluxes over a wide spectral range (from 0.33 up to 12~$\mu$m) that we obtained combining HST, 2MASS \citep{Skrutskie2006}, and WISE data \citep{Wright2010}. The HST magnitudes were calibrated into flux using the recipes in the WFC3 handbook\footnote{ https://hst-docs.stsci.edu/display/WFC3IHB}.



Figure~\ref{fig:sed} shows a comparison between  the SED of GQ~Lup~C and a BT-Settl model spectrum \citep{Allard2012} with the parameters closest to those found in Paper I, i.e. T$_{\rm eff}$=3200 K and $\log{g} =3.5$~dex. An extinction of $A_V\sim $1.0~mag, as found in Paper I, has been also considered. The agreement is excellent in the spectral range between 0.4 and 2.2~$\mu$m. However, the HST data shows evidence for strong excesses in the H$\alpha$ narrow band filter and in the UV, that are clear signs of mass accretion from the disk. By comparing the flux in the F656N filter with respect to that measured in broad band filters, and taking into account the band width, we obtained an equivalent width of $188\pm 3$~\AA\ for H$\alpha$, that is considerably stronger than the value of 100~\AA\ obtained from the X-Shooter spectrum in Paper I. This argues for a strong variability of the source, which has been frequently observed in accreting objects (see, e.g., \citealt{Frasca2018}, and references therein).

The SED also shows an IR excess at wavelengths longer than 3~$\mu$m, consistent with the detection of the warm part of the disk around GQ~Lup~C. This may be fitted by a black body SED of $800\pm 100$~K, a value that is constrained at the upper edge by the observed K magnitude, and at the lower edge by the non detection in W3. This is lower than the dust sublimation temperature, suggesting a real truncation of the disk. More specifically, this temperature is about 1/4 of the stellar temperature, suggesting that the disk is truncated at $R_{\rm T}\sim 15-20~R_*$, $R_*$ being the stellar radius. This may be compared with the value of  $R_{\rm T}/R_*$ given by eq. (2.2) of \citet{Bouvier2007}: assuming B=1~kG, $\dot{M}=3.3\times 10^{-10}$~M$_\odot$/yr, $M_*=0.15$~M$_\odot$ and $R_*=0.87$~R$\odot$ (Paper I), we obtain $R_{\rm T}/R*$=12.3. Integrating the fluxes from the model atmosphere and the black body representing the emission from the warm disk over all wavelengths, we obtain a value of $L_{\rm disk} = L_{\rm bol} - L_{\rm star} = 0.053~L_{\rm star}$, where $L_{\rm bol}$ is the total luminosity; this is of course a lower limit to the disk emission because we cannot exclude that there is also a cold component emitting at wavelengths longer than those considered here. Using Spitzer data to estimate $L_{\rm bol}$, \citet{Merin2008} obtained an average value of $L_{\rm disk}=0.25~L_{\rm star}$, but with a considerable scatter with values ranging from 0.01 to 1.5. Usually, accretors are thought to have $L_{\rm disk}/L_{\rm star}> 0.08$. GQ~Lup~C would then be at the lower edge of the distribution of this quantity.

\section{Disk images from HST}

There is a star (Gaia DR2 source 6011522757637551616) with similar peak intensity less than 2 arcsec from GQ~Lup~C (farther from GQ Lup A). This star can be used as a comparison source for HST images because it is placed at similar separation from the field center and it is quite bright. A comparison of the flux of GQ~Lup~C with that of this comparison star confirms the clear detection of a strong H$\alpha$\ emission and UV excess for GQ~Lup~C. In addition and most interestingly, GQ~Lup~C image appears to be fuzzier in comparison with the comparison star. While peak intensity is lower at most wavelengths, the FWHM of the images are consistently larger (about 130 mas with respect to about 80 mas for the comparison object). This suggests that the disk around GQ~Lup~C is actually resolved from the stellar PSF in the HST images. 

\begin{figure}
   \centering
   \includegraphics[width=\columnwidth]{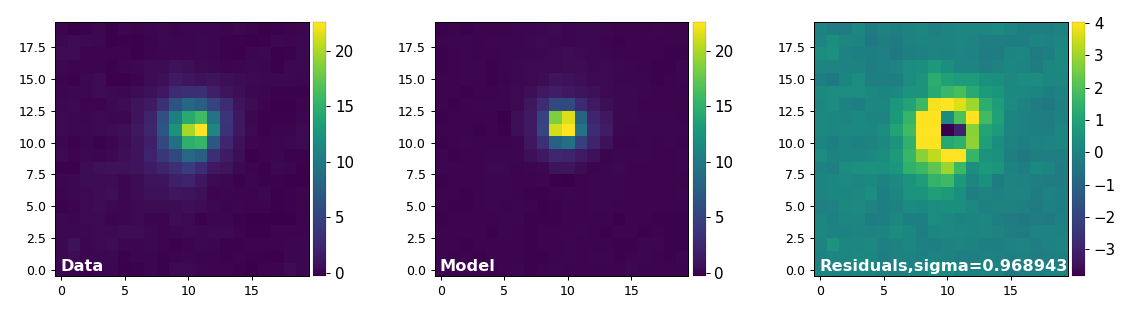}
   \caption{The disk of GQ~Lup~C from the HST data after application of the PSF subtraction procedure. In this figure, North is up and East to the left and scale is in pixel (one pixel equal to 0.03962 arcsec)}
   \label{fig:disk}
   \end{figure}

In order to image the disk from the HST data we used a technique based on the PSF subtraction (see Appendix A for a more extended discussion). As a first step, in each frame we determined the position and the flux for GQ~Lup~C. Then we considered the PSF of the reference star and rescaled it to the flux obtained for GQ~Lup~C. As a last step, we subtracted the reference PSF to our target star in its position and rotated the frame with respect to the True North (see next Section). The residuals are shown in the last panel of Figure \ref{fig:disk}, whereas the first two images show GQ~Lup~C and the model PSF used for the subtraction, respectively. 


The PSF subtraction yields a well resolved image of the disk of GQ~Lup~C. We consider here the image obtained by combining the four best bands (F850LP, F625W, F555W, F775W); however, very similar results are obtained for the individual bands. After PSF subtraction, the excess signal looks roughly distributed along an ellipse aligned NW-SE, with the E side brighter than the W one; the central empty region is likely an artefact of the procedure. We think this light distribution is produced by scattering of the stellar light by a flared circular disk seen at rather high inclination, with the near side less luminous because very close to the center of the star image and then subtracted by the procedure. 

   \begin{figure}[htb]
   \centering
   \includegraphics[width=\columnwidth]{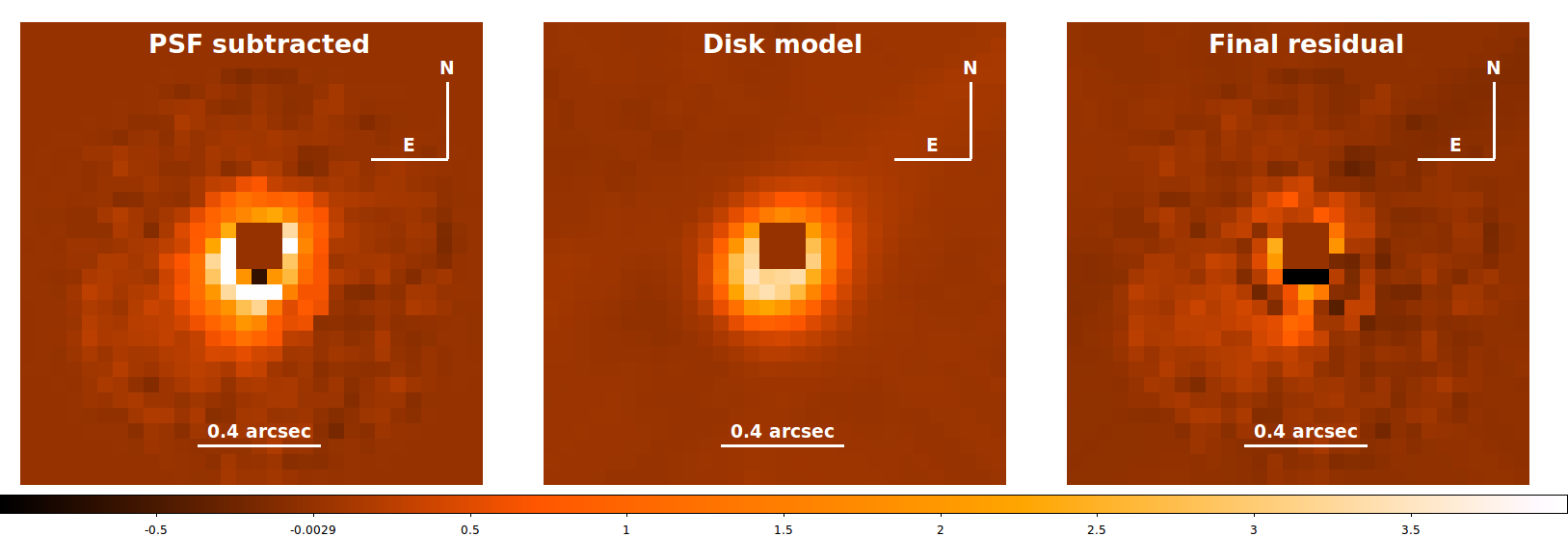}
   \caption{Illustration of the method used for the estimation of the disk parameters. Left panel: image after PSF subtraction; central panel: the model disk; right panel: same as the left panel, but after subtraction of the model disk}
   \label{fig:disk_sub}
   \end{figure}

In order to determine the geometrical parameters of this disk, we assumed that the disk is actually a circular ring of radius $r$\ seen at a given inclination $i$\ and position angle $PA$, and convoluted it with the PSF of the nearby comparison star. We considered that the ring can be offset with respect to the center of the star along the minor axis of the projected ellipse, as seen in numerous cases of disks seen in scattered light (see e.g. \citealt{Ginski2016, Sissa2018}). We then found the set of parameters that minimizes the r.m.s. of the residuals between the image obtained after the PSF subtraction and the model disk. Figure~\ref{fig:disk_sub} illustrates the procedure. We repeated this procedure for the images obtained with the individual bands and assumed the mean value of the parameters as best estimate, and the standard deviation of the mean as its uncertainty. In this way, we obtained a radius of $r=2.22\pm 0.14$~pixels (that is $88\pm 6$~mas, or $13.2\pm 0.8$~au at the distance of GQ~Lup), an inclination $i=44\pm 2$~degree, a position angle $PA=315\pm 4$~degree, and an offset of $0.97\pm 0.15$~pixel (that is, $38\pm 6$~mas) in the NE direction. The offset can be interpreted as due to flaring of the disk at an angle of about $26\pm 7$~degree, the near side being SW.

For comparison, the disk of GQ~Lup~A has $PA=346\pm 1$\ and $i=60.5\pm 0.5$~degree, as derived from ALMA data \citep{MacGregor2017}; similar parameters ($PA=349\pm 5$ and $i=56\pm 5$) were obtained by \citet{Wu2017}. In addition, if we assume that the circumprimary disk seen by ALMA is co-rotating with GQ~Lup~B, that is describing a counter-clockwise orbit (see \citealt{Schwarz2016}), the near side is the SW one also in that case. This comparison suggests that the two disks have a similar though not identical orientation, being disaligned by $38.1\pm 3.4$~degree. 

\section{Is there an outflow from GQ~Lup~C?}

Astrometric calibration of the HST images was obtained using as reference the position, proper motion and parallax of GQ~Lup A, C, and of two background stars (sources id. 6011522757637561216 and 6011522757637551616) measured by Gaia DR2 \citep{Gaia2016, Gaia2018}. In this way, we obtained a plate scale of 0.03962 arcsec/pixel (very close to the nominal value of 0.0395 arcsec/pixel given in the WFC3 handbook\footnote{https://hst-docs.stsci.edu/display/WFC3IHB} and a true north correction of 132.153 degree.

We then measured the position of GQ~Lup~C in the individual bands with respect to the two reference background stars, and between the two reference stars themselves. We found that the position in most bands are essentially identical with each other, with an r.m.s. scatter of about 6~mas. However, the position of GQ~Lup~C measured on the image obtained with the F656N filter (that is, at H$\alpha$) is offset with respect to the average of that obtained at the remaining wavelengths by about 13~mas, at $PA=306\pm 24$~degree measured on sky. While this result has a significance at only slightly more than twice the error bar, it suggests that H$\alpha$ is emitted from a region that is not the same as for the remaining wavelengths. This also agrees with the smaller value of the FWHM of the GQ~Lup~C image (112 mas) obtained with F656N with respect an average value of 131 mas (with an r.m.s. of 10 mas) for the remaining wavelengths.

We propose that most of the H$\alpha$ emission may be attributed to an outflow from GQ~Lup~C. The outflow is expected to be seen projected along the minor axis of the disk image on the far side, that is roughly at PA=45 degree. This implies that the H$\alpha$ emission seems to be shifted at almost right angle, but slightly in the opposite direction. This last feature might be explained as follows: while the direct star light is heavily absorbed by the disk and we see it largely as light scattered by the disk - that is, at a position substantially offset with respect to the real position of the star along the minor axis of the disk -, the outflow is not significantly obscured. This agrees well with the discussion in Paper I. Variability of the emission is consistent with the small size of the outflow ($\sim 0.1$~arcsec) and with the quite long time elapsed between HST and X-Shooter observations (seven years).

\section{Discussion and Conclusions}

The main result of this letter is the detection of the disk around GQ~Lup~C both from the excess in the WISE W1 and W2 bands and from scattered light seen in the HST images. The near-infrared excess reveals the warm part of the disk, whose emission is matched by a black body with a temperature of 800~K. We notice that this temperature is lower than that of dust sublimation, suggesting that the disk is actually truncated, probably by magnetic fields at the stellar co-rotation radius. This is not well determined, because we can only estimate the stellar rotation period from the values of the stellar radius and of $V \sin{i}$ determined in Paper I, and using the inclination $i$\ we obtained for the disk. The corresponding period is 2.7~d, but it is largely uncertain because of the huge error bar of $V \sin{i}$. On the other hand, the inner disk radius is also uncertain because the value of the albedo is not well determined. However, an inner disk radius of 0.20~au, corresponding to a period of 2.7~d for the stellar mass (0.15~M$_\odot$) determined in Paper I is obtained assuming an albedo of 0.3.

The scattered light image provides an estimate of the disk radius of $13.2\pm 0.8$~au. This is smaller than all disk radii obtained in the ALMA survey of disks in Lupus \citep{Ansdell2016, Ansdell2018}; however this is due to the resolution of the ALMA images used in that survey, because there are examples of disks as small as that in GQ~Lup~C resolved by ALMA (see e.g. \citealt{Facchini2019}). On the other hand, this same survey suggests the presence of a correlation between disk radii and dust masses, that is roughly described by the power law $M_{\rm dust}/M_\oplus \sim 0.044~(R_{\rm dust}/au)^{1.5}$. While this relation is based on sub-mm observations, that can provide very different images of disks with respect to scattered light, we notice that if we use the disk radius determined from the HST data, we would obtain a dust mass of 2.1~M$_\oplus$; this is a quite typical value for disks around small mass stars in Lupus \citep{Ansdell2016, Ansdell2018}. The disk of GQ~Lup~C looks then not anomalous for the Lupus association, and we expect that the corresponding sub-mm flux would have been detected by that survey, though resolution of the disk would have required a longer baseline.

We finally note that the orientation of the major axis of the disks of GQ Lup A and C agrees fairly well with the main axis of the filament visible in Herschel images \citep{Rygl2013}, while that of the minor axis agrees with that of the magnetic field in nearby regions \citep{Rizzo1998, Matthews2014}. 

\begin{acknowledgements}
R.G., D.M., S.D. and V.D. acknowledge support from the ``Progetti Premiali" funding scheme of the Italian Ministry of Education, University, and Research. This work has been supported by the project PRIN-INAF 2016 The Cradle of Life - GENESIS-SKA (General Conditions in Early Planetary Systems for the rise of life with SKA). E.R. is supported by the European Union's Horizon 2020 research and innovation programme under the Marie Sk\l odowska-Curie grant agreement No 664931. This work has made use of data from the European Space Agency (ESA) mission {\it Gaia} (\url{https://www.cosmos.esa.int/gaia}), processed by the {\it Gaia} Data Processing and Analysis Consortium (DPAC, \url{https://www.cosmos.esa.int/web/gaia/dpac/consortium}). Funding for the DPAC has been provided by national institutions, in particular the institutions participating in the {\it Gaia} Multilateral Agreement.
\end{acknowledgements}

\bibliographystyle{aa} 
\bibliography{main.bib} %

\begin{thebibliography}{20}
\expandafter\ifx\csname natexlab\endcsname\relax\def\natexlab#1{#1}\fi

\bibitem[{{Alcal{\'a}} {et~al.}(2020){Alcal{\'a}}, {Majidi}, {Desidera},
  {Frasca}, {Manara}, {Rigliaco}, {Gratton}, {Bonnefoy}, {Covino}, {Chauvin},
  {Claudi}, {D'Orazi}, {Langlois}, {Lazzoni}, {Mesa}, {Schlieder}, \&
  {Vigan}}]{Alcala2020}
{Alcal{\'a}}, J.~M., {Majidi}, F.~Z., {Desidera}, S., {et~al.} 2020, arXiv
  e-prints, arXiv:2001.10879

\bibitem[{{Allard} {et~al.}(2012){Allard}, {Homeier}, \&
  {Freytag}}]{Allard2012}
{Allard}, F., {Homeier}, D., \& {Freytag}, B. 2012, Philosophical Transactions
  of the Royal Society of London Series A, 370, 2765

\bibitem[{{Ansdell} {et~al.}(2018){Ansdell}, {Williams}, {Trapman}, {van
  Terwisga}, {Facchini}, {Manara}, {van der Marel}, {Miotello}, {Tazzari},
  {Hogerheijde}, {Guidi}, {Testi}, \& {van Dishoeck}}]{Ansdell2018}
{Ansdell}, M., {Williams}, J.~P., {Trapman}, L., {et~al.} 2018, \apj, 859, 21

\bibitem[{{Ansdell} {et~al.}(2016){Ansdell}, {Williams}, {van der Marel},
  {Carpenter}, {Guidi}, {Hogerheijde}, {Mathews}, {Manara}, {Miotello},
  {Natta}, {Oliveira}, {Tazzari}, {Testi}, {van Dishoeck}, \& {van
  Terwisga}}]{Ansdell2016}
{Ansdell}, M., {Williams}, J.~P., {van der Marel}, N., {et~al.} 2016, \apj,
  828, 46

\bibitem[{{Bouvier} {et~al.}(2007){Bouvier}, {Alencar}, {Harries},
  {Johns-Krull}, \& {Romanova}}]{Bouvier2007}
{Bouvier}, J., {Alencar}, S.~H.~P., {Harries}, T.~J., {Johns-Krull}, C.~M., \&
  {Romanova}, M.~M. 2007, in Protostars and Planets V, ed. B.~{Reipurth},
  D.~{Jewitt}, \& K.~{Keil}, 479

\bibitem[{{Facchini} {et~al.}(2019){Facchini}, {van Dishoeck}, {Manara},
  {Tazzari}, {Maud}, {Cazzoletti}, {Rosotti}, {van der Marel}, {Pinilla}, \&
  {Clarke}}]{Facchini2019}
{Facchini}, S., {van Dishoeck}, E.~F., {Manara}, C.~F., {et~al.} 2019, \aap,
  626, L2

\bibitem[{{Frasca} {et~al.}(2018){Frasca}, {Montes}, {Alcal{\`a}}, {Klutsch},
  \& {Guillout}}]{Frasca2018}
{Frasca}, A., {Montes}, D., {Alcal{\`a}}, J.~M., {Klutsch}, A., \& {Guillout},
  P. 2018, \actaa, 68, 403

\bibitem[{{Gaia Collaboration} {et~al.}(2018){Gaia Collaboration}, {Brown},
  {Vallenari}, {Prusti}, {de Bruijne}, {Babusiaux}, {Bailer-Jones}, {Biermann},
  {Evans}, {Eyer}, {Jansen}, {Jordi}, {Klioner}, {Lammers}, {Lindegren},
  {Luri}, {Mignard}, {Panem}, {Pourbaix}, {Randich}, {Sartoretti}, {Siddiqui},
  {Soubiran}, {van Leeuwen}, {Walton}, {Arenou}, {Bastian}, {Cropper},
  {Drimmel}, {Katz}, {Lattanzi}, {Bakker}, {Cacciari}, {Casta{\~n}eda},
  {Chaoul}, {Cheek}, {De Angeli}, {Fabricius}, {Guerra}, {Holl}, {Masana},
  {Messineo}, {Mowlavi}, {Nienartowicz}, {Panuzzo}, {Portell}, {Riello},
  {Seabroke}, {Tanga}, {Th{\'e}venin}, {Gracia-Abril}, {Comoretto},
  {Garcia-Reinaldos}, {Teyssier}, {Altmann}, {Andrae}, {Audard},
  {Bellas-Velidis}, {Benson}, {Berthier}, {Blomme}, {Burgess}, {Busso},
  {Carry}, {Cellino}, {Clementini}, {Clotet}, {Creevey}, {Davidson}, {De
  Ridder}, {Delchambre}, {Dell'Oro}, {Ducourant},
  {Fern{\'a}ndez-Hern{\'a}ndez}, {Fouesneau}, {Fr{\'e}mat}, {Galluccio},
  {Garc{\'\i}a-Torres}, {Gonz{\'a}lez-N{\'u}{\~n}ez}, {Gonz{\'a}lez-Vidal},
  {Gosset}, {Guy}, {Halbwachs}, {Hambly}, {Harrison}, {Hern{\'a}ndez},
  {Hestroffer}, {Hodgkin}, {Hutton}, {Jasniewicz}, {Jean-Antoine-Piccolo},
  {Jordan}, {Korn}, {Krone-Martins}, {Lanzafame}, {Lebzelter}, {L{\"o}ffler},
  {Manteiga}, {Marrese}, {Mart{\'\i}n-Fleitas}, {Moitinho}, {Mora}, {Muinonen},
  {Osinde}, {Pancino}, {Pauwels}, {Petit}, {Recio-Blanco}, {Richards},
  {Rimoldini}, {Robin}, {Sarro}, {Siopis}, {Smith}, {Sozzetti}, {S{\"u}veges},
  {Torra}, {van Reeven}, {Abbas}, {Abreu Aramburu}, {Accart}, {Aerts},
  {Altavilla}, {{\'A}lvarez}, {Alvarez}, {Alves}, {Anderson}, {Andrei},
  {Anglada Varela}, {Antiche}, {Antoja}, {Arcay}, {Astraatmadja}, {Bach},
  {Baker}, {Balaguer-N{\'u}{\~n}ez}, {Balm}, {Barache}, {Barata}, {Barbato},
  {Barblan}, {Barklem}, {Barrado}, {Barros}, {Barstow}, {Bartholom{\'e}
  Mu{\~n}oz}, {Bassilana}, {Becciani}, {Bellazzini}, {Berihuete}, {Bertone},
  {Bianchi}, {Bienaym{\'e}}, {Blanco-Cuaresma}, {Boch}, {Boeche}, {Bombrun},
  {Borrachero}, {Bossini}, {Bouquillon}, {Bourda}, {Bragaglia}, {Bramante},
  {Breddels}, {Bressan}, {Brouillet}, {Br{\"u}semeister}, {Brugaletta},
  {Bucciarelli}, {Burlacu}, {Busonero}, {Butkevich}, {Buzzi}, {Caffau},
  {Cancelliere}, {Cannizzaro}, {Cantat-Gaudin}, {Carballo}, {Carlucci},
  {Carrasco}, {Casamiquela}, {Castellani}, {Castro-Ginard}, {Charlot},
  {Chemin}, {Chiavassa}, {Cocozza}, {Costigan}, {Cowell}, {Crifo}, {Crosta},
  {Crowley}, {Cuypers}, {Dafonte}, {Damerdji}, {Dapergolas}, {David}, {David},
  {de Laverny}, {De Luise}, {De March}, {de Martino}, {de Souza}, {de Torres},
  {Debosscher}, {del Pozo}, {Delbo}, {Delgado}, {Delgado}, {Di Matteo},
  {Diakite}, {Diener}, {Distefano}, {Dolding}, {Drazinos}, {Dur{\'a}n},
  {Edvardsson}, {Enke}, {Eriksson}, {Esquej}, {Eynard Bontemps}, {Fabre},
  {Fabrizio}, {Faigler}, {Falc{\~a}o}, {Farr{\`a}s Casas}, {Federici},
  {Fedorets}, {Fernique}, {Figueras}, {Filippi}, {Findeisen}, {Fonti},
  {Fraile}, {Fraser}, {Fr{\'e}zouls}, {Gai}, {Galleti}, {Garabato},
  {Garc{\'\i}a-Sedano}, {Garofalo}, {Garralda}, {Gavel}, {Gavras}, {Gerssen},
  {Geyer}, {Giacobbe}, {Gilmore}, {Girona}, {Giuffrida}, {Glass}, {Gomes},
  {Granvik}, {Gueguen}, {Guerrier}, {Guiraud}, {Guti{\'e}rrez-S{\'a}nchez},
  {Haigron}, {Hatzidimitriou}, {Hauser}, {Haywood}, {Heiter}, {Helmi}, {Heu},
  {Hilger}, {Hobbs}, {Hofmann}, {Holland}, {Huckle}, {Hypki}, {Icardi},
  {Jan{\ss}en}, {Jevardat de Fombelle}, {Jonker}, {Juh{\'a}sz}, {Julbe},
  {Karampelas}, {Kewley}, {Klar}, {Kochoska}, {Kohley}, {Kolenberg},
  {Kontizas}, {Kontizas}, {Koposov}, {Kordopatis}, {Kostrzewa-Rutkowska},
  {Koubsky}, {Lambert}, {Lanza}, {Lasne}, {Lavigne}, {Le Fustec}, {Le
  Poncin-Lafitte}, {Lebreton}, {Leccia}, {Leclerc}, {Lecoeur-Taibi},
  {Lenhardt}, {Leroux}, {Liao}, {Licata}, {Lindstr{\o}m}, {Lister}, {Livanou},
  {Lobel}, {L{\'o}pez}, {Managau}, {Mann}, {Mantelet}, {Marchal}, {Marchant},
  {Marconi}, {Marinoni}, {Marschalk{\'o}}, {Marshall}, {Martino}, {Marton},
  {Mary}, {Massari}, {Matijevi{\v{c}}}, {Mazeh}, {McMillan}, {Messina},
  {Michalik}, {Millar}, {Molina}, {Molinaro}, {Moln{\'a}r}, {Montegriffo},
  {Mor}, {Morbidelli}, {Morel}, {Morris}, {Mulone}, {Muraveva}, {Musella},
  {Nelemans}, {Nicastro}, {Noval}, {O'Mullane}, {Ord{\'e}novic},
  {Ord{\'o}{\~n}ez-Blanco}, {Osborne}, {Pagani}, {Pagano}, {Pailler},
  {Palacin}, {Palaversa}, {Panahi}, {Pawlak}, {Piersimoni}, {Pineau}, {Plachy},
  {Plum}, {Poggio}, {Poujoulet}, {Pr{\v{s}}a}, {Pulone}, {Racero}, {Ragaini},
  {Rambaux}, {Ramos-Lerate}, {Regibo}, {Reyl{\'e}}, {Riclet}, {Ripepi}, {Riva},
  {Rivard}, {Rixon}, {Roegiers}, {Roelens}, {Romero-G{\'o}mez}, {Rowell},
  {Royer}, {Ruiz-Dern}, {Sadowski}, {Sagrist{\`a} Sell{\'e}s}, {Sahlmann},
  {Salgado}, {Salguero}, {Sanna}, {Santana-Ros}, {Sarasso}, {Savietto},
  {Schultheis}, {Sciacca}, {Segol}, {Segovia}, {S{\'e}gransan}, {Shih},
  {Siltala}, {Silva}, {Smart}, {Smith}, {Solano}, {Solitro}, {Sordo}, {Soria
  Nieto}, {Souchay}, {Spagna}, {Spoto}, {Stampa}, {Steele},
  {Steidelm{\"u}ller}, {Stephenson}, {Stoev}, {Suess}, {Surdej}, {Szabados},
  {Szegedi-Elek}, {Tapiador}, {Taris}, {Tauran}, {Taylor}, {Teixeira},
  {Terrett}, {Teyssand ier}, {Thuillot}, {Titarenko}, {Torra Clotet}, {Turon},
  {Ulla}, {Utrilla}, {Uzzi}, {Vaillant}, {Valentini}, {Valette}, {van Elteren},
  {Van Hemelryck}, {van Leeuwen}, {Vaschetto}, {Vecchiato}, {Veljanoski},
  {Viala}, {Vicente}, {Vogt}, {von Essen}, {Voss}, {Votruba}, {Voutsinas},
  {Walmsley}, {Weiler}, {Wertz}, {Wevers}, {Wyrzykowski}, {Yoldas},
  {{\v{Z}}erjal}, {Ziaeepour}, {Zorec}, {Zschocke}, {Zucker}, {Zurbach}, \&
  {Zwitter}}]{Gaia2018}
{Gaia Collaboration}, {Brown}, A.~G.~A., {Vallenari}, A., {et~al.} 2018, \aap,
  616, A1

\bibitem[{{Gaia Collaboration} {et~al.}(2016){Gaia Collaboration}, {Prusti},
  {de Bruijne}, {Brown}, {Vallenari}, {Babusiaux}, {Bailer-Jones}, {Bastian},
  {Biermann}, {Evans}, {Eyer}, {Jansen}, {Jordi}, {Klioner}, {Lammers},
  {Lindegren}, {Luri}, {Mignard}, {Milligan}, {Panem}, {Poinsignon},
  {Pourbaix}, {Randich}, {Sarri}, {Sartoretti}, {Siddiqui}, {Soubiran},
  {Valette}, {van Leeuwen}, {Walton}, {Aerts}, {Arenou}, {Cropper}, {Drimmel},
  {H{\o}g}, {Katz}, {Lattanzi}, {O'Mullane}, {Grebel}, {Holland}, {Huc},
  {Passot}, {Bramante}, {Cacciari}, {Casta{\~n}eda}, {Chaoul}, {Cheek}, {De
  Angeli}, {Fabricius}, {Guerra}, {Hern{\'a}ndez}, {Jean-Antoine-Piccolo},
  {Masana}, {Messineo}, {Mowlavi}, {Nienartowicz}, {Ord{\'o}{\~n}ez-Blanco},
  {Panuzzo}, {Portell}, {Richards}, {Riello}, {Seabroke}, {Tanga},
  {Th{\'e}venin}, {Torra}, {Els}, {Gracia-Abril}, {Comoretto},
  {Garcia-Reinaldos}, {Lock}, {Mercier}, {Altmann}, {Andrae}, {Astraatmadja},
  {Bellas-Velidis}, {Benson}, {Berthier}, {Blomme}, {Busso}, {Carry},
  {Cellino}, {Clementini}, {Cowell}, {Creevey}, {Cuypers}, {Davidson}, {De
  Ridder}, {de Torres}, {Delchambre}, {Dell'Oro}, {Ducourant}, {Fr{\'e}mat},
  {Garc{\'\i}a-Torres}, {Gosset}, {Halbwachs}, {Hambly}, {Harrison}, {Hauser},
  {Hestroffer}, {Hodgkin}, {Huckle}, {Hutton}, {Jasniewicz}, {Jordan},
  {Kontizas}, {Korn}, {Lanzafame}, {Manteiga}, {Moitinho}, {Muinonen},
  {Osinde}, {Pancino}, {Pauwels}, {Petit}, {Recio-Blanco}, {Robin}, {Sarro},
  {Siopis}, {Smith}, {Smith}, {Sozzetti}, {Thuillot}, {van Reeven}, {Viala},
  {Abbas}, {Abreu Aramburu}, {Accart}, {Aguado}, {Allan}, {Allasia},
  {Altavilla}, {{\'A}lvarez}, {Alves}, {Anderson}, {Andrei}, {Anglada Varela},
  {Antiche}, {Antoja}, {Ant{\'o}n}, {Arcay}, {Atzei}, {Ayache}, {Bach},
  {Baker}, {Balaguer-N{\'u}{\~n}ez}, {Barache}, {Barata}, {Barbier}, {Barblan},
  {Baroni}, {Barrado y Navascu{\'e}s}, {Barros}, {Barstow}, {Becciani},
  {Bellazzini}, {Bellei}, {Bello Garc{\'\i}a}, {Belokurov}, {Bendjoya},
  {Berihuete}, {Bianchi}, {Bienaym{\'e}}, {Billebaud}, {Blagorodnova},
  {Blanco-Cuaresma}, {Boch}, {Bombrun}, {Borrachero}, {Bouquillon}, {Bourda},
  {Bouy}, {Bragaglia}, {Breddels}, {Brouillet}, {Br{\"u}semeister},
  {Bucciarelli}, {Budnik}, {Burgess}, {Burgon}, {Burlacu}, {Busonero}, {Buzzi},
  {Caffau}, {Cambras}, {Campbell}, {Cancelliere}, {Cantat-Gaudin}, {Carlucci},
  {Carrasco}, {Castellani}, {Charlot}, {Charnas}, {Charvet}, {Chassat},
  {Chiavassa}, {Clotet}, {Cocozza}, {Collins}, {Collins}, {Costigan}, {Crifo},
  {Cross}, {Crosta}, {Crowley}, {Dafonte}, {Damerdji}, {Dapergolas}, {David},
  {David}, {De Cat}, {de Felice}, {de Laverny}, {De Luise}, {De March}, {de
  Martino}, {de Souza}, {Debosscher}, {del Pozo}, {Delbo}, {Delgado},
  {Delgado}, {di Marco}, {Di Matteo}, {Diakite}, {Distefano}, {Dolding}, {Dos
  Anjos}, {Drazinos}, {Dur{\'a}n}, {Dzigan}, {Ecale}, {Edvardsson}, {Enke},
  {Erdmann}, {Escolar}, {Espina}, {Evans}, {Eynard Bontemps}, {Fabre},
  {Fabrizio}, {Faigler}, {Falc{\~a}o}, {Farr{\`a}s Casas}, {Faye}, {Federici},
  {Fedorets}, {Fern{\'a}ndez-Hern{\'a}ndez}, {Fernique}, {Fienga}, {Figueras},
  {Filippi}, {Findeisen}, {Fonti}, {Fouesneau}, {Fraile}, {Fraser}, {Fuchs},
  {Furnell}, {Gai}, {Galleti}, {Galluccio}, {Garabato}, {Garc{\'\i}a-Sedano},
  {Gar{\'e}}, {Garofalo}, {Garralda}, {Gavras}, {Gerssen}, {Geyer}, {Gilmore},
  {Girona}, {Giuffrida}, {Gomes}, {Gonz{\'a}lez-Marcos},
  {Gonz{\'a}lez-N{\'u}{\~n}ez}, {Gonz{\'a}lez-Vidal}, {Granvik}, {Guerrier},
  {Guillout}, {Guiraud}, {G{\'u}rpide}, {Guti{\'e}rrez-S{\'a}nchez}, {Guy},
  {Haigron}, {Hatzidimitriou}, {Haywood}, {Heiter}, {Helmi}, {Hobbs},
  {Hofmann}, {Holl}, {Holland }, {Hunt}, {Hypki}, {Icardi}, {Irwin}, {Jevardat
  de Fombelle}, {Jofr{\'e}}, {Jonker}, {Jorissen}, {Julbe}, {Karampelas},
  {Kochoska}, {Kohley}, {Kolenberg}, {Kontizas}, {Koposov}, {Kordopatis},
  {Koubsky}, {Kowalczyk}, {Krone-Martins}, {Kudryashova}, {Kull}, {Bachchan},
  {Lacoste-Seris}, {Lanza}, {Lavigne}, {Le Poncin-Lafitte}, {Lebreton},
  {Lebzelter}, {Leccia}, {Leclerc}, {Lecoeur-Taibi}, {Lemaitre}, {Lenhardt},
  {Leroux}, {Liao}, {Licata}, {Lindstr{\o}m}, {Lister}, {Livanou}, {Lobel},
  {L{\"o}ffler}, {L{\'o}pez}, {Lopez-Lozano}, {Lorenz}, {Loureiro},
  {MacDonald}, {Magalh{\~a}es Fernandes}, {Managau}, {Mann}, {Mantelet},
  {Marchal}, {Marchant}, {Marconi}, {Marie}, {Marinoni}, {Marrese},
  {Marschalk{\'o}}, {Marshall}, {Mart{\'\i}n-Fleitas}, {Martino}, {Mary},
  {Matijevi{\v{c}}}, {Mazeh}, {McMillan}, {Messina}, {Mestre}, {Michalik},
  {Millar}, {Miranda}, {Molina}, {Molinaro}, {Molinaro}, {Moln{\'a}r},
  {Moniez}, {Montegriffo}, {Monteiro}, {Mor}, {Mora}, {Morbidelli}, {Morel},
  {Morgenthaler}, {Morley}, {Morris}, {Mulone}, {Muraveva}, {Musella},
  {Narbonne}, {Nelemans}, {Nicastro}, {Noval}, {Ord{\'e}novic},
  {Ordieres-Mer{\'e}}, {Osborne}, {Pagani}, {Pagano}, {Pailler}, {Palacin},
  {Palaversa}, {Parsons}, {Paulsen}, {Pecoraro}, {Pedrosa}, {Pentik{\"a}inen},
  {Pereira}, {Pichon}, {Piersimoni}, {Pineau}, {Plachy}, {Plum}, {Poujoulet},
  {Pr{\v{s}}a}, {Pulone}, {Ragaini}, {Rago}, {Rambaux}, {Ramos-Lerate},
  {Ranalli}, {Rauw}, {Read}, {Regibo}, {Renk}, {Reyl{\'e}}, {Ribeiro},
  {Rimoldini}, {Ripepi}, {Riva}, {Rixon}, {Roelens}, {Romero-G{\'o}mez},
  {Rowell}, {Royer}, {Rudolph}, {Ruiz-Dern}, {Sadowski}, {Sagrist{\`a}
  Sell{\'e}s}, {Sahlmann}, {Salgado}, {Salguero}, {Sarasso}, {Savietto},
  {Schnorhk}, {Schultheis}, {Sciacca}, {Segol}, {Segovia}, {Segransan},
  {Serpell}, {Shih}, {Smareglia}, {Smart}, {Smith}, {Solano}, {Solitro},
  {Sordo}, {Soria Nieto}, {Souchay}, {Spagna}, {Spoto}, {Stampa}, {Steele},
  {Steidelm{\"u}ller}, {Stephenson}, {Stoev}, {Suess}, {S{\"u}veges}, {Surdej},
  {Szabados}, {Szegedi-Elek}, {Tapiador}, {Taris}, {Tauran}, {Taylor},
  {Teixeira}, {Terrett}, {Tingley}, {Trager}, {Turon}, {Ulla}, {Utrilla},
  {Valentini}, {van Elteren}, {Van Hemelryck}, {van Leeuwen}, {Varadi},
  {Vecchiato}, {Veljanoski}, {Via}, {Vicente}, {Vogt}, {Voss}, {Votruba},
  {Voutsinas}, {Walmsley}, {Weiler}, {Weingrill}, {Werner}, {Wevers},
  {Whitehead}, {Wyrzykowski}, {Yoldas}, {{\v{Z}}erjal}, {Zucker}, {Zurbach},
  {Zwitter}, {Alecu}, {Allen}, {Allende Prieto}, {Amorim},
  {Anglada-Escud{\'e}}, {Arsenijevic}, {Azaz}, {Balm}, {Beck}, {Bernstein},
  {Bigot}, {Bijaoui}, {Blasco}, {Bonfigli}, {Bono}, {Boudreault}, {Bressan},
  {Brown}, {Brunet}, {Bunclark}, {Buonanno}, {Butkevich}, {Carret}, {Carrion},
  {Chemin}, {Ch{\'e}reau}, {Corcione}, {Darmigny}, {de Boer}, {de Teodoro}, {de
  Zeeuw}, {Delle Luche}, {Domingues}, {Dubath}, {Fodor}, {Fr{\'e}zouls},
  {Fries}, {Fustes}, {Fyfe}, {Gallardo}, {Gallegos}, {Gardiol}, {Gebran},
  {Gomboc}, {G{\'o}mez}, {Grux}, {Gueguen}, {Heyrovsky}, {Hoar}, {Iannicola},
  {Isasi Parache}, {Janotto}, {Joliet}, {Jonckheere}, {Keil}, {Kim},
  {Klagyivik}, {Klar}, {Knude}, {Kochukhov}, {Kolka}, {Kos}, {Kutka}, {Lainey},
  {LeBouquin}, {Liu}, {Loreggia}, {Makarov}, {Marseille}, {Martayan},
  {Martinez-Rubi}, {Massart}, {Meynadier}, {Mignot}, {Munari}, {Nguyen},
  {Nordlander}, {Ocvirk}, {O'Flaherty}, {Olias Sanz}, {Ortiz}, {Osorio},
  {Oszkiewicz}, {Ouzounis}, {Palmer}, {Park}, {Pasquato}, {Peltzer}, {Peralta},
  {P{\'e}turaud}, {Pieniluoma}, {Pigozzi}, {Poels}, {Prat}, {Prod'homme},
  {Raison}, {Rebordao}, {Risquez}, {Rocca-Volmerange}, {Rosen}, {Ruiz-Fuertes},
  {Russo}, {Sembay}, {Serraller Vizcaino}, {Short}, {Siebert}, {Silva},
  {Sinachopoulos}, {Slezak}, {Soffel}, {Sosnowska}, {Strai{\v{z}}ys}, {ter
  Linden}, {Terrell}, {Theil}, {Tiede}, {Troisi}, {Tsalmantza}, {Tur},
  {Vaccari}, {Vachier}, {Valles}, {Van Hamme}, {Veltz}, {Virtanen}, {Wallut},
  {Wichmann}, {Wilkinson}, {Ziaeepour}, \& {Zschocke}}]{Gaia2016}
{Gaia Collaboration}, {Prusti}, T., {de Bruijne}, J.~H.~J., {et~al.} 2016,
  \aap, 595, A1

\bibitem[{{Ginski} {et~al.}(2016){Ginski}, {Stolker}, {Pinilla}, {Dominik},
  {Boccaletti}, {de Boer}, {Benisty}, {Biller}, {Feldt}, {Garufi}, {Keller},
  {Kenworthy}, {Maire}, {M{\'e}nard}, {Mesa}, {Milli}, {Min}, {Pinte}, {Quanz},
  {van Boekel}, {Bonnefoy}, {Chauvin}, {Desidera}, {Gratton}, {Girard},
  {Keppler}, {Kopytova}, {Lagrange}, {Langlois}, {Rouan}, \&
  {Vigan}}]{Ginski2016}
{Ginski}, C., {Stolker}, T., {Pinilla}, P., {et~al.} 2016, \aap, 595, A112

\bibitem[{{MacGregor} {et~al.}(2017){MacGregor}, {Wilner}, {Czekala},
  {Andrews}, {Dai}, {Herczeg}, {Kratter}, {Kraus}, {Ricci}, \&
  {Testi}}]{MacGregor2017}
{MacGregor}, M.~A., {Wilner}, D.~J., {Czekala}, I., {et~al.} 2017, \apj, 835,
  17

\bibitem[{{Matthews} {et~al.}(2014){Matthews}, {Ade}, {Angil{\`e}}, {Benton},
  {Chapin}, {Chapman}, {Devlin}, {Fissel}, {Fukui}, {Gandilo}, {Gundersen},
  {Hargrave}, {Klein}, {Korotkov}, {Moncelsi}, {Mroczkowski}, {Netterfield},
  {Novak}, {Nutter}, {Olmi}, {Pascale}, {Poidevin}, {Savini}, {Scott},
  {Shariff}, {Soler}, {Tachihara}, {Thomas}, {Truch}, {Tucker}, {Tucker}, \&
  {Ward-Thompson}}]{Matthews2014}
{Matthews}, T.~G., {Ade}, P. A.~R., {Angil{\`e}}, F.~E., {et~al.} 2014, \apj,
  784, 116

\bibitem[{{Mer{\'\i}n} {et~al.}(2008){Mer{\'\i}n}, {J{\o}rgensen}, {Spezzi},
  {Alcal{\'a}}, {Evans}, {Harvey}, {Prusti}, {Chapman}, {Huard}, {van
  Dishoeck}, \& {Comer{\'o}n}}]{Merin2008}
{Mer{\'\i}n}, B., {J{\o}rgensen}, J., {Spezzi}, L., {et~al.} 2008, \apjs, 177,
  551

\bibitem[{{Rizzo} {et~al.}(1998){Rizzo}, {Morras}, \& {Arnal}}]{Rizzo1998}
{Rizzo}, J.~R., {Morras}, R., \& {Arnal}, E.~M. 1998, \mnras, 300, 497

\bibitem[{{Rygl} {et~al.}(2013){Rygl}, {Benedettini}, {Schisano}, {Elia},
  {Molinari}, {Pezzuto}, {Andr{\'e}}, {Bernard}, {White}, {Polychroni},
  {Bontemps}, {Cox}, {Di Francesco}, {Facchini}, {Fallscheer}, {di Giorgio},
  {Hennemann}, {Hill}, {K{\"o}nyves}, {Minier}, {Motte}, {Nguyen-Luong},
  {Peretto}, {Pestalozzi}, {Sadavoy}, {Schneider}, {Spinoglio}, {Testi}, \&
  {Ward-Thompson}}]{Rygl2013}
{Rygl}, K.~L.~J., {Benedettini}, M., {Schisano}, E., {et~al.} 2013, \aap, 549,
  L1

\bibitem[{{Schwarz} {et~al.}(2016){Schwarz}, {Ginski}, {de Kok}, {Snellen},
  {Brogi}, \& {Birkby}}]{Schwarz2016}
{Schwarz}, H., {Ginski}, C., {de Kok}, R.~J., {et~al.} 2016, \aap, 593, A74

\bibitem[{{Sissa} {et~al.}(2018){Sissa}, {Gratton}, {Garufi}, {Rigliaco},
  {Zurlo}, {Mesa}, {Langlois}, {de Boer}, {Desidera}, {Ginski}, {Lagrange},
  {Maire}, {Vigan}, {Dima}, {Antichi}, {Baruffolo}, {Bazzon}, {Benisty},
  {Beuzit}, {Biller}, {Boccaletti}, {Bonavita}, {Bonnefoy}, {Brandner},
  {Bruno}, {Buenzli}, {Cascone}, {Chauvin}, {Cheetham}, {Claudi}, {Cudel}, {De
  Caprio}, {Dominik}, {Fantinel}, {Farisato}, {Feldt}, {Fontanive}, {Galicher},
  {Giro}, {Hagelberg}, {Incorvaia}, {Janson}, {Kasper}, {Keppler}, {Kopytova},
  {Lagadec}, {Lannier}, {Lazzoni}, {LeCoroller}, {Lessio}, {Ligi}, {Marzari},
  {Menard}, {Meyer}, {Mouillet}, {Peretti}, {Perrot}, {Potiron}, {Rouan},
  {Salasnich}, {Salter}, {Samland}, {Schmidt}, {Scuderi}, \&
  {Wildi}}]{Sissa2018}
{Sissa}, E., {Gratton}, R., {Garufi}, A., {et~al.} 2018, \aap, 619, A160

\bibitem[{{Skrutskie} {et~al.}(2006){Skrutskie}, {Cutri}, {Stiening},
  {Weinberg}, {Schneider}, {Carpenter}, {Beichman}, {Capps}, {Chester},
  {Elias}, {Huchra}, {Liebert}, {Lonsdale}, {Monet}, {Price}, {Seitzer},
  {Jarrett}, {Kirkpatrick}, {Gizis}, {Howard}, {Evans}, {Fowler}, {Fullmer},
  {Hurt}, {Light}, {Kopan}, {Marsh}, {McCallon}, {Tam}, {Van Dyk}, \&
  {Wheelock}}]{Skrutskie2006}
{Skrutskie}, M.~F., {Cutri}, R.~M., {Stiening}, R., {et~al.} 2006, \aj, 131,
  1163

\bibitem[{{Wright} {et~al.}(2010){Wright}, {Eisenhardt}, {Mainzer}, {Ressler},
  {Cutri}, {Jarrett}, {Kirkpatrick}, {Padgett}, {McMillan}, {Skrutskie},
  {Stanford}, {Cohen}, {Walker}, {Mather}, {Leisawitz}, {Gautier}, {McLean},
  {Benford}, {Lonsdale}, {Blain}, {Mendez}, {Irace}, {Duval}, {Liu}, {Royer},
  {Heinrichsen}, {Howard}, {Shannon}, {Kendall}, {Walsh}, {Larsen}, {Cardon},
  {Schick}, {Schwalm}, {Abid}, {Fabinsky}, {Naes}, \& {Tsai}}]{Wright2010}
{Wright}, E.~L., {Eisenhardt}, P. R.~M., {Mainzer}, A.~K., {et~al.} 2010, \aj,
  140, 1868

\bibitem[{{Wu} {et~al.}(2017){Wu}, {Sheehan}, {Males}, {Close}, {Morzinski},
  {Teske}, {Haug-Baltzell}, {Merchant}, \& {Lyons}}]{Wu2017}
{Wu}, Y.-L., {Sheehan}, P.~D., {Males}, J.~R., {et~al.} 2017, \apj, 836, 223

\end{thebibliography}

\begin{appendix}
\section{PSF subtraction method}
The method applied in Section 4 to image the disk around GQ Lup C after the subtraction of a model PSF is described in details in Lazzoni et al. (submitted). Here, we just summarize two main caveats to be considered when using this technique. The first one regards the model used, since we are fitting the parameters of a Gaussian profile for the central peak and damped correction for the Airy diffraction pattern (the model PSF of the nearby star) to a PSF that has a different shape (GQ Lup C), given by the presence of the disk. For this reason, the optimized residuals will show over-bright areas surrounded by over-subtracted ones to balance the incongruous models. Thus, one of the most difficult parameters to estimate is the flux of the source whereas the geometry of the disk should be less affected, especially for centered disks around the star. In cases like the one presented in this paper, where the disk looks slightly mis-centered, the derived inclination and width are to be considered as a first estimate only.\\
The second caveat regards, instead, the differences between the PSF of the object and the one of the model due to the instruments itself. Indeed, even if the two PSFs are taken at the same time (thus in the same conditions), distortions may emerge from different position in the field of view. However, since the two sources are quite close on the detector, we can neglect such effects for this case. 
\section{Tables}
\begin{table}[h]
\caption[]{Data for spectral energy distribution of GQ~Lup~C}
\label{tab:sed}
\begin{tabular}{lccc}
\hline
Band & Wavel. & Magnitude & Flux \\
     & ($\mu$m) & (mag) & erg/cm$^2$/s$^1$/nm \\ 
\hline     
HST-F336W  & 0.3355 & $21.631\pm 0.046$ & $(7.21\pm 0.30)$E-17 \\
HST-F390W  & 0.3921 & $22.258\pm 0.108$ & $(7.23\pm 0.68)$E-17 \\
HST-F475W  & 0.4771 & $21.052\pm 0.018$ & $(1.97\pm 0.03)$E-16 \\
HST-F555W  & 0.5305 & $20.362\pm 0.018$ & $(2.82\pm 0.05)$E-16 \\
HST-F625W  & 0.6241 & $19.106\pm 0.017$ & $(5.52\pm 0.09)$E-16 \\
HST-F656N  & 0.6561 & $16.086\pm 0.015$ & $(5.24\pm 0.07)$E-16 \\
HST-F673N  & 0.6766 & $18.992\pm 0.020$ & $(4.80\pm 0.09)$E-16 \\
HST-F775W  & 0.7651 & $17.487\pm 0.011$ & $(1.32\pm 0.01)$E-15 \\
HST-F850LP & 0.9187 & $16.366\pm 0.012$ & $(2.25\pm 0.02)$E-15 \\
2MASS-J    & 1.235  & $14.849\pm 0.052$ & $(3.60\pm 0.17)$E-15 \\
2MASS-H    & 1.662  & $14.083\pm 0.048$ & $(2.64\pm 0.11)$E-15 \\
2MASS-K    & 2.159  & $13.818\pm 0.049$ & $(1.27\pm 0.06)$E-15 \\
WISE-W1    & 3.353  &  $12.60\pm 0.38$  & $(7.47\pm 2.21)$E-16 \\
WISE-W2    & 4.603  &  $12.20\pm 0.40$  & $(3.17\pm 0.98)$E-16 \\
WISE-W3    & 12     &    $<$10.638      &$<$3.62E-17\\
\hline     
\end{tabular}
\end{table}

\end{appendix}

\end{document}